\documentclass[11pt,a4paper]{article}
\usepackage[T1]{fontenc}
\usepackage{amsmath,amsfonts, amssymb, amsbsy, mathrsfs, mathtools, bbm}
\usepackage{titlesec}
\usepackage{enumitem}
\usepackage{float}
\usepackage{authblk}
\usepackage{makeidx}
\usepackage{graphicx}
\usepackage{wrapfig}
\usepackage{subfig}
\usepackage{color,mdframed}
\usepackage{mathpazo}
\usepackage[toc]{appendix}
\usepackage[hidelinks,colorlinks=true,linkcolor=red,citecolor=blue,urlcolor=magenta,breaklinks=true]{hyperref}
\usepackage[left=2.00cm, right=2.0cm, top=3.0cm, bottom=3.00cm]{geometry}

\newcommand{\email}[1]{\href{mailto:#1}{#1}}

\newcommand{\df}{\textrm{d}}
\newcommand{\cg}{\textnormal{\textsl{g}}}

\newcommand{\ex}{\textrm{e}}
\newcommand{\hf}{{\frac{1}{2}}}

\newcommand{\la}{\left\langle}
\newcommand{\ra}{\right\rangle}

\newcommand{\io}{{\textrm{i}}}

\newcommand{\bz}{\boldsymbol{Z}}
\newcommand{\bc}{\boldsymbol{C}}
\newcommand{\Br}{\mathcal{B}}

\newcommand{\tmg}{^{\rm (MG)}T}

\titleformat{\paragraph}
{\normalfont\normalsize\bfseries}{\theparagraph}{1em}{}
\titlespacing*{\paragraph}
{0pt}{3.25ex plus 1ex minus .2ex}{1.5ex plus .2ex}

\titleformat{\subparagraph}
{\normalfont\normalsize\bfseries}{\thesubparagraph}{1em}{}
\titlespacing*{\subparagraph}
{0pt}{3.25ex plus 1ex minus .2ex}{1.5ex plus .2ex}

\makeatletter
\def\blfootnote{\xdef\@thefnmark{}\@footnotetext}
\makeatother

\usepackage[square, numbers, sort&compress]{natbib}

\numberwithin{equation}{section}

\begin{document}
	\title{\textbf{An Almost-FLRW Universe as the Averaged Geometry in Macroscopic Gravity}}
	\author{Anish Agashe\footnote{\email{anish.agashe@utdallas.edu}}~}
	\author{Mustapha Ishak\footnote{\email{mishak@utdallas.edu}}}
	\affil{\normalsize Department of Physics, The University of Texas at Dallas, Richardson, TX 75080, USA}
	
	\date{}
	\maketitle
	\begin{abstract}
		It is well-known that spacetime averaging is an operation that does not commute with building the Einstein tensor. In the framework of Macroscopic gravity (MG), a covariant averaging procedure, this non-commutativity gives averaged field equations with an additional correction term known as back-reaction. It is important to explore whether such a term, even if known to be small, may or may not cause any systematic effect for precision cosmology. In this work, we explore the application of the MG formalism to an almost Friedmann-Lema\^itre-Robertson-Walker (FLRW) model. Namely, we find solutions to the field equations of MG taking the averaged universe to be almost-FLRW modelled using a linearly perturbed FLRW metric. We study several solutions with different functional forms of the metric perturbations including plane waves ansatzes. We find that back-reaction terms are present not only at the background level but also at perturbed level, reflecting the non-linear nature of the averaging process. Thus, the averaging effect can extend to both the expansion and the growth of structure in the universe.
	\end{abstract}
	
	\maketitle
	
	\section{Introduction}
	When building a cosmological model in the framework of general relativity (GR), an effective averaging of the real discrete matter distribution is carried out on the right hand side of the Einstein field equations (EFEs), while the left hand side -- the Einstein tensor -- is assumed to describe the gravitational behaviour of this matter distribution without being explicitly averaged \cite{shiro1,shiro2,shiro3}. Though this has been a standard practice in cosmology, it has been put into question \cite{ellis3,tava}. An obvious approach to deal with this issue would be to take an explicit average of the EFEs. Indeed, a number of authors have argued for the use of field equations that are averaged \cite{ellis2,ellis3,ellis4,tava,shiro1,shiro2,shiro3}.
	
	The construction of such averaged field equations, however, is not straightforward. It is difficult to define a mathematically rigorous formulation to calculate covariant volume averages of tensor fields on the pseudo-Riemannian space-time of general relativity. The difficulties are exacerbated due to the non-linear nature of Einstein field equations. This means that the operation of taking the average and calculating Einstein tensor do not commute. In other words, the average of Einstein tensor for a microscopic geometry is not equal to the Einstein tensor for the averaged (macroscopic) geometry. It was suggested in \cite{shiro1,ellis2}, and has been well accepted since (see \cite{cliftonbr,ellisbr1,hoogen1} and the references therein), that any form of averaged field equations of gravity built from EFEs should have a correction term that will account for this non-commutativity. This correction, arising in the macroscopic description of gravity, is referred to as back-reaction. The need for an averaged theory of gravity; the difficulties in formulating such a theory; and given such a theory, the task of calculating and interpreting the back-reaction, together constitute what is known as the averaging problem in general relativity and cosmology.
	
	There is a long history of approaches taken towards solving this problem \cite{ellis2,ellis3,ellis4,tava,shiro1,brill,szek,noo1,noo2,boersma,zoto1,zoto2,stoeger,kasai1,kasai2,bild,fut,fut1,fut2,fut3,fut4,fut5,carfora,nambu,wilt1,wilt2,wilt3,wilt4,brannlund,hoogen,hoogen3,russ1,zala1,zala2,zala3,zala4,zala5,zala6,zala7,buch1,buch2,buch3,buch5,buch6,buch7,buch8,buch9,buch10,wald2,lca1,lca2,lca3,lca4,lca5,lca6}. With various approaches at hand, some debate has ensued as well. A few years back, there was a debate regarding the relevance of the back-reaction for cosmology between Green and Wald, and Buchert et. al \cite{buchertbr1}. Green and Wald argued that back-reaction is irrelevant for the large scale dynamics (see \cite{wald2} and subsequent work), however, Buchert et al. criticised their formalism to be not general enough. Green and Wald put forth a rebuttal to this \cite{wald6} but did point out that their definition of back-reaction is different from that of the other approaches. Recently, the two averaging approaches and their relation to space-time averaging was compared for simple exact space-times (specifically, spherically symmetric and plane symmetric space-times) in \cite{clifton1}. Let us clarify that in this work, we do not take sides in this debate and it is outside the scope of our work to comment on the validity of any of the approaches involved. It seems to us that even if back-reaction cannot have large dynamical effects, it remains to be determined either from theory or observations if back-reaction can have any percent or sub-percent level contributions to observational cosmology like other astrophysical systematics. In this work, we push the framework of macroscopic gravity further to make it comparable to observations (as done in other previous papers \cite{clarkson,clifton,tharake1,tharake2}) by analysing almost-FLRW models within macroscopic gravity. 
	
	Macroscopic Gravity (MG) \cite{zala1,zala2,zala3,zala4,zala5,zala6,zala7} is a covariant and exact (non-perturbative) approach to calculate volume averages of arbitrary tensor fields on an $n$-dimensional differentiable manifold. The averages are calculated by the virtue of bilocal averaging operators and Lie dragging the averaging regions \cite{zala1,zala2,zala3,zala4,zala5,zala6,zala7}. Applying this procedure to the $4$-dimensional pseudo-Riemannian geometry of GR yields the macroscopic Einstein field equations (mEFEs). In these equations, the effect of averaging (the back-reaction) is characterised by a tensorial correction term, $\bc$. The averaged field equations can be written in the form of EFEs by taking this correction term on the right hand side of the averaged equations. Then, this term can be regarded as a geometrical correction to the averaged matter-energy distribution \cite{zala1,zala2,zala3,zala4,zala5,zala6,zala7}. 
	
	This correction term is made of contractions of a tensor called the connection correlation tensor, $\bz$. The connection correlation tensor satisfies (by construction) certain algebraic and differential conditions. By imposing constraints on the structure of $\bz$ and assuming a macroscopic geometry, these conditions can be solved to find the value of the correction term $\bc$ in terms of the components of $\bz$ and the (assumed) metric coefficients. This has been done for exact geometries like FLRW  \cite{coley,clifton,hoogen2}, Bianchi Type-I \cite{tharake1} and static spherically symmetric (Schwarzschild) \cite{hoogenss}. In \cite{coleyss1,coleyss2}, the authors took the microscopic geometry to be Lema\^itre-Tolman-Bondi (LTB) and averaged it using the MG procedure. In \cite{clifton,para1,para2,para3,para4}, perturbations around the exact-FLRW solutions were considered. The FLRW solution in MG was constrained using observational data in \cite{clarkson,tharake1,tharake2} and it was found that the back-reaction is small and is only significant at the level of sub-percent systematics in precision cosmology.    
	
	In this paper, we find solutions to the field equations in MG taking the averaged universe to be almost-FLRW. We consider that after applying the averaging procedure of MG to the small scale lumpy universe, we reach a scale where the inhomogeneities are almost smoothed out, but not completely. We calculate the MG correction term for this scale. It has indeed been argued in \cite{ellisbr,ellis5,mattsson,buch4} that the averaging process should be multi-scale. That is, the effect of averaging at any scale should be the cumulative effect of all the coarse-graining done to reach the smoothness at that scale. The almost-FLRW geometry is conventionally modelled by taking perturbations around the exact-FLRW universe. To this end, we solve the field equations of MG with the macroscopic geometry given by a linearly perturbed FLRW metric. The back-reaction at this scale can then be considered to be due to all the higher order perturbations that have been averaged over. 
	
	Linear perturbations around FLRW background in the context of MG have been previously analysed in \cite{clifton,para1,para2,para3,para4,tharake1,tharake2}. There, it was assumed that $\bz$ can be split into zeroth, $^{(0)}\bz$, and first order, $^{(1)}\bz$, terms. Then, the constraint equations for $^{(1)}\bz$  \cite{tharake1} were solved separately. In this work, we take a rather passive approach. We do not split $\bz$ beforehand. Instead, considering the averaged geometry of the universe to be linearly perturbed FLRW, we explicitly calculate $\bz$ in terms of the metric coefficients just like one calculates other quantities (Einstein tensor, connection, etc.) in standard cosmological perturbation theory.
	
	We restrict ourselves to scalar perturbations and find several solutions with various assumptions on their functional form and space/time dependencies. We find that when considered as arbitrary functions, only severely restricted form of inhomogeneous perturbations lead to non-trivial back-reaction terms. However, more general solutions can be found by taking an ansatz on the functional form of the perturbations. We find that $\bc$ can be split in terms that are zeroth and first order in perturbations. Therefore, when working with almost-FLRW geometry, the back-reaction has a component that enters the field equations at first order. That is, the back-reaction modifies the evolution of not just the background, but also the perturbations around it. We also find that, in some particular cases, the back-reaction terms can lead to a non-zero slip parameter even in the absence of anisotropic stress in the matter-energy content of the universe.
	
	The paper is arranged as follows: in section \ref{mgtheory}, we present an overview of macroscopic gravity averaging scheme and describe the constraints on the connection correlation tensor. Then, in section \ref{mgsol}, we explain how to solve these constraints in a systematic manner to find the correction tensor $\bc$ in terms of the  metric coefficients. Then, we present our results for the MG correction term with an almost-FLRW metric in section \ref{almostflrw}. Readers familiar with the MG theory and its solutions can skip directly to this section. We calculate the correction term for plane wave perturbations in section \ref{pwa}. In section \ref{secpertmefe}, we present the full field equations for an almost-FLRW metric. Section \ref{conclusion} summarises the results in this paper with some remarks.
	
	The notation and convention used is as follows: tensors associated with the microscopic geometry are denoted by lowercase letters and those with the macroscopic geometry by uppercase letters. Greek indices run from $0$ to $3$ and Latin indices from $1$ to $3$. Angular brackets $ \la\cdots\ra $ denote the averaging operation or sometimes averaged quantities. Covariant differentiation with respect to the macroscopic connection is denoted by $\|$. Indices with round brackets $ (~) $/square brackets $[~]$ are symmetrised/anti-symmetrised; and  underlined indices are not included in (anti-)symmetrisation. The sign convention followed is the `Landau-Lifshitz Space-like Convention (LLSC)' \cite{mtw}. That is, the signature of the metric is taken to be Lorentzian $ (-,+,+,+) $, the Riemann curvature tensor is defined as, $ {r^\mu}_{\alpha\nu\beta}~=~2\partial_{[\nu}{\gamma^\mu}_{\underline{\alpha}\beta]} + 2{\gamma^\epsilon}_{\alpha[\beta}{\gamma^\mu}_{\underline{\epsilon}\nu]} $ and $ r_{\mu\nu}~=~{r^\alpha}_{\mu\alpha\nu} $ is the Ricci tensor. The Ricci scalar is defined as $ r~=~{r^\mu}_\mu~=~\cg^{\mu\nu}r_{\mu\nu} $.  Finally, the units are taken such that $ G=1=c $, i.e., $ \kappa = 8\pi $.   
	
	\section{Macroscopic Gravity Formalism}\label{mgtheory}
	
	Using the concepts of macroscopic electrodynamics \cite{russ,lorentz,jackson}, a covariant averaging procedure was introduced in \cite{zala1,zala2,zala3,zala4,zala5,zala6,zala7} which can be used in general relativity. It is a generalisation of the averaging in Minkowski space-time and is valid for arbitrary classical tensor fields on any differentiable manifold \cite{zala1,zala2,zala3,zala4,zala5,zala6,zala7}. Applying this procedure to the $4$-dimensional pseudo-Riemannian geometry of GR leads to an averaged theory of gravity -- Macroscopic Gravity.
	
	Let there be a geometric object, $ p^\alpha_\beta(x) $, defined on an $ n $-dimensional differentiable metric manifold $ (\mathcal{M}, \cg_{\alpha\beta}) $. Then, the space-time averaged value of this object over a compact region $ \Sigma \subset \mathcal{M} $ with a volume $ n $-form around a supporting point $ x \in \Sigma $, is defined as,
	\begin{equation} \label{mgavg}
		\la p^\alpha_\beta(x) \ra = \dfrac{\int_\Sigma \mathcal{A}^\alpha_{\mu^\prime}(x,x^\prime) p^{\mu^\prime}_{\nu^\prime}(x^\prime) \mathcal{A}^{\nu^\prime}_\beta(x^\prime, x) \sqrt{-\cg^\prime}\ \df^n x^\prime}{\int_\Sigma \sqrt{-\cg^\prime}\ \df^n x^\prime}
	\end{equation}
	where $ \int_\Sigma \sqrt{-\cg^\prime}\ \df^n x^\prime $ is the volume ($ V_\Sigma $) of the region $ \Sigma $ and $ \cg^\prime = \det\left[\cg_{\alpha\beta}(x^\prime)\right] $. The integration is done over all the points $ x^\prime \in \Sigma $. The integrand $ \mathcal{A}^\alpha_{\mu^\prime}(x,x^\prime) p^{\mu^\prime}_{\nu^\prime}(x^\prime) \mathcal{A}^{\nu^\prime}_\beta(x^\prime, x) $ is called the bilocal extension of the object $ p^\alpha_\beta(x) $; and the objects $ \mathcal{A}^\alpha_{\mu^\prime}(x,x^\prime) $ and $ \mathcal{A}^{\nu^\prime}_\beta(x^\prime, x) $ are called the bilocal averaging operators. In the next subsections, we describe come of the underlying concepts and definition in macroscopic gravity
	
	\subsection{Basic Definitions}
	The average of the microscopic Riemann curvature tensor, $ \la{r^\alpha}_{\beta\rho\sigma}\ra $ is written as $ {R^\alpha}_{\beta\rho\sigma} $ and it is assumed to behave like a curvature tensor itself \cite{zala1,zala2,zala3,zala4,zala5,zala6,zala7}. It can be shown to satisfy the algebraic and differential properties of the Riemann curvature tensor. Further, bilocal objects defined as, $ {\mathcal{F}^\alpha}_{\beta\rho}(x,x^\prime)\equiv \mathcal{A}^\alpha_{\epsilon^\prime}\left(\partial_\rho\mathcal{A}^{\epsilon^\prime}_\beta + \nabla_{\sigma^\prime}\mathcal{A}^{\epsilon^\prime}_\beta\mathcal{A}^{\sigma^\prime}_\rho\right) $, behave as connection coefficients at $ x $ and hence, can be considered to be the bilocal extension of the microscopic connection coefficients, $ {\gamma^\alpha}_{\beta\rho} $. They have the following coincidence limit,
	\begin{equation}\label{Fcoinlim}
		\lim_{x^\prime\to x} {\mathcal{F}^\alpha}_{\beta\rho}  =  {\gamma^\alpha}_{\beta\rho} 
	\end{equation} 
	
	The average of these objects, $ {\overline{\mathcal{F}}^\alpha}_{\beta\rho}\equiv\frac{\int {\mathcal{F}^\alpha}_{\beta\rho} \sqrt{-\cg^\prime}\ \df^n x^\prime}{\int \sqrt{-\cg^\prime}\ \df^n x^\prime}=\la{\mathcal{F}^\alpha}_{\beta\rho}\ra$, is interpreted as the affine connection coefficients of the averaged space-time \cite{zala1,zala2}. The curvature tensor, $ {M^\alpha}_{\beta\rho\sigma} $, associated with these connection coefficients is the ``macroscopic'' curvature tensor, written as,
	\begin{equation}\label{macrocurv}
		{M^\alpha}_{\beta\rho\sigma} = {R^\alpha}_{\beta\rho\sigma} + 2\la{\mathcal{F}^\delta}_{\beta[\rho}{\mathcal{F}^\alpha}_{\underline{\delta}\sigma]}\ra - 2\la{\mathcal{F}^\delta}_{\beta[\rho}\ra\la{\mathcal{F}^\alpha}_{\underline{\delta}\sigma]}\ra
	\end{equation}
	where $ {R^\alpha}_{\beta\rho\sigma} = \la {r^\alpha}_{\beta\rho\sigma} \ra $ is the average of the microscopic Riemann curvature tensor.
	
	One can calculate another connection, $ {\Pi^\alpha}_{\beta\rho} $, associated with the curvature tensor $ {R^\alpha}_{\beta\rho\sigma} $. Therefore, we have two connections associated with two curvature tensors which are related to each other by equation \eqref{macrocurv}. The difference between these two connections is captured by another object, namely, the affine deformation tensor, $ {A^\alpha}_{\beta\rho} = \la{\mathcal{F}^\alpha}_{\beta\rho}\ra - {\Pi^\alpha}_{\beta\rho}$. This is a consequence of the non-metricity of $ {R^\alpha}_{\beta\rho\sigma} $, i.e., the average of microscopic metric, $ \la\cg_{\alpha\beta}\ra $, is not a metric tensor, and,	
	\begin{equation}\label{nonmet1}
		G_{\alpha\beta\lvert\rho} \ne 0
	\end{equation}
	where, $ G_{\alpha\beta} $ is the metric tensor associated with the averaged manifold, and $ \lvert $ denotes the covariant derivative with respect to the connection $ {\Pi^\alpha}_{\beta\rho} $. That is, we have a macroscopic metric, $ G_{\alpha\beta} $, associated with the macroscopic space-time which is not compatible with the non-Riemannian curvature, $ {R^\alpha}_{\beta\rho\sigma} $, and its associated connection $ {\Pi^\alpha}_{\beta\rho} $, while the average of the microscopic metric tensor, $ \la\cg_{\alpha\beta}\ra $, is not the metric associated with the macroscopic space-time, and is not a metric tensor.
	
	Further, looking at the construction of both the curvature tensors in terms of their respective connections, equation \eqref{macrocurv} takes the form,
	\begin{equation}\label{curvdef}
		{A^\alpha}_{\beta[\rho\|\sigma]} - {A^\epsilon}_{\beta[\rho}{A^\alpha}_{\underline{\epsilon}\sigma]} = -\hf{Q^\alpha}_{\beta\rho\sigma}
	\end{equation}
	where, 
	\begin{subequations}
		\begin{equation}\label{curvdef1}
			{Q^\alpha}_{\beta\rho\sigma} = 2\la{\mathcal{F}^\delta}_{\beta[\rho}{\mathcal{F}^\alpha}_{\underline{\delta}\sigma]}\ra - 2\la{\mathcal{F}^\delta}_{\beta[\rho}\ra\la{\mathcal{F}^\alpha}_{\underline{\delta}\sigma]}\ra
		\end{equation}	
		\textrm{\indent The tensor, $ {Q^\alpha}_{\beta\rho\sigma} $, in its mathematical construction, is like a curvature tensor, built from the affine deformation tensor $ {A^\alpha}_{\beta\rho} $ in the same way $ {M^\alpha}_{\beta\rho\sigma} $ is built from its associated connection. From equations \eqref{macrocurv} and \eqref{curvdef1}, we can see the that it measures the difference between the two curvature tensors -- a curvature deformation tensor,}
		\begin{equation}\label{curvdef2}
			{Q^\alpha}_{\beta\rho\sigma} = {M^\alpha}_{\beta\rho\sigma} - {R^\alpha}_{\beta\rho\sigma} 
		\end{equation}
	\end{subequations}
	Both the tensors, $ {Q^\alpha}_{\beta\rho\sigma} $ and $ {M^\alpha}_{\beta\rho\sigma} $, follow the algebraic properties of a Riemann curvature tensor. The Ricci tensors $ M_{\alpha\beta} = {M^\epsilon}_{\alpha\epsilon\beta} $ and $ R_{\alpha\beta} = {R^\epsilon}_{\alpha\epsilon\beta} $ are symmetric. The Ricci tensor equivalent for $ {Q^\alpha}_{\beta\rho\sigma} $ can be written as $ Q_{\alpha\beta} = {Q^\epsilon}_{\alpha\epsilon\beta} $.
	
	Moreover, both the curvature tensors of the averaged geometry satisfy the differential Bianchi identities with respect to their corresponding connections. The averaging of the differential Bianchi identities for the microscopic curvature tensor gives the following relations,
	\begin{equation}\label{bianchiavg}
		{R^\alpha}_{\beta[\rho\sigma,\epsilon]} + \la{r^\gamma}_{\beta[\rho\sigma}{\mathcal{F}^\alpha}_{\underline{\gamma}\epsilon]}\ra - \la{r^\alpha}_{\gamma[\rho\sigma}{\mathcal{F}^\gamma}_{\underline{\beta}\epsilon]}\ra = 0
	\end{equation}	
	To solve the above equations completely, one needs a splitting rule for the averaging operation present in the second and third terms. Such a rule can be constructed by using an object called the connection correlation, which is defined as \cite{zala1,zala2},
	\begin{equation}\label{corr2form}
		{Z^\alpha}_{\beta[\gamma}{^\mu}_{\underline{\nu}\sigma]} = \la {\mathcal{F}^\alpha}_{\beta[\gamma}{\mathcal{F}^\mu}_{\underline{\nu}\sigma]} \ra - \la {\mathcal{F}^\alpha}_{\beta[\gamma}\ra \la{\mathcal{F}^\mu}_{\underline{\nu}\sigma]} \ra
	\end{equation} 
	
	Using equations \eqref{curvdef1} and \eqref{corr2form}, we see that,
	\begin{equation}\label{curvdefcorr2form}
		{Q^\alpha}_{\beta\rho\sigma} = 2{Z^\epsilon}_{\beta[\rho}{^\alpha}_{\underline{\epsilon}\sigma]}
	\end{equation}
	
	Therefore, to summarise, when the averaging procedure in equation \eqref{mgavg} is applied to the 4-dimensional Riemannian manifold in general relativity, the averaged manifold is characterised by a metric, two equi-affine symmetric connections, two curvature tensors associated with these connections and correlation tensors constructed to solve differential Bianchi identities for these curvature tensors. The number of correlation tensors is finite and restricted by the dimensionality of the manifold.
	
	\subsection{The Macroscopic Einstein Field Equations}
	The microscopic Einstein field equations are given by,
	\begin{equation}\label{microefe}
		{e^\epsilon}_\gamma = {r^\epsilon}_\gamma - \hf\delta^\epsilon_\gamma \cg^{\mu\nu}r_{\mu\nu} = \kappa {t^\epsilon}_\gamma		
	\end{equation}
	where, ${e^\epsilon}_\gamma$ is the Einstein tensor\footnote{We use $\boldsymbol{e}/\boldsymbol{E}$ to denote the Einstein tensor, instead of $\boldsymbol{g}/\boldsymbol{G}$, to avoid confusion with the metric tensor.} and $ {r^\epsilon}_\gamma = \cg^{\alpha\epsilon}r_{\alpha\gamma} $ is the Ricci tensor. The matter-energy content of the microscopic manifold is described by the microscopic energy-momentum tensor $ {t^\epsilon}_\gamma $. Then, the average of Einstein field equations,
	\begin{equation}\label{avgefe}
		\la {e^\epsilon}_\gamma \ra = \la \cg^{\beta\epsilon}r_{\beta\gamma}\ra - \hf\delta^\epsilon_\gamma\la \cg^{\mu\nu}r_{\mu\nu}\ra = \kappa\la {t^\epsilon}_\gamma\ra
	\end{equation}
	yields the macroscopic Einstein field equations (mEFEs), which are given by,
	\begin{equation}\label{macroefe}
		{E^\epsilon}_\gamma = \la \cg^{\beta\epsilon}\ra M_{\beta\gamma} - \hf\delta^\epsilon_\gamma\la \cg^{\mu\nu}\ra M_{\mu\nu} = \kappa\la {t^\epsilon}_\gamma\ra + \left({Z^\epsilon}_{\mu\nu\gamma} - \hf\delta^\epsilon_\gamma Q_{\mu\nu}\right)\la \cg^{\mu\nu}\ra 
	\end{equation}
	where $ {Z^\epsilon}_{\mu\nu\gamma} = 2{Z^\epsilon}_{\mu[\alpha}{^\alpha}_{\underline{\nu}\gamma]} $ is a Ricci-tensor-like object for the connection correlation and $ Q_{\mu\nu} = {Q^\epsilon}_{\mu\epsilon\nu} $. Therefore, in the theory of macroscopic gravity, averaging out the Einstein field equations introduces additional terms constituting various traces (contractions) of the connection correlation.	
	
	The appearance of these additional terms in equation \eqref{macroefe} can be considered as a correction arising from the change in the geometric structure of the space-time due to averaging. We will call this term, the MG correction term, and denote it as,
	\begin{equation}\label{tmg}
		{C^\epsilon}_\gamma = \left({Z^\epsilon}_{\mu\nu\gamma} - \frac{1}{2}\delta^\epsilon_\gamma Q_{\mu\nu}\right)\la \cg^{\mu\nu}\ra
	\end{equation}
	Further, we can introduce a macroscopic energy-momentum tensor, $ {T^\epsilon}_\gamma $, which is defined as,
	\begin{equation}\label{macroem}
		{T^\epsilon}_\gamma = \la {t^\epsilon}_\gamma\ra + {\tmg^\epsilon}_\gamma
	\end{equation}
	where $ {\tmg^\epsilon}_\gamma = \frac{1}{\kappa}{C^\epsilon}_\gamma $. 
	Using this definition, the mEFEs take the following elegant form,
	\begin{equation}\label{macroefe1}
		\la \cg^{\beta\epsilon}\ra M_{\beta\gamma} - \hf\delta^\epsilon_\gamma\la \cg^{\mu\nu}\ra M_{\mu\nu} = \kappa {T^\epsilon}_\gamma
	\end{equation} 
	
	\subsection{Macroscopic Gravity Objects}
	\subsubsection{The Connection Correlation Tensor} \label{zconstraints}
	The connection correlation tensor is an important object that exists on the averaged space-time and has the following algebraic and differential properties \cite{zala1,zala2,zala3,zala4,zala5,zala6,zala7},
	\begin{subequations}\label{corr2formprop}			
		\begin{align}
			\intertext{The anti-symmetry in the third and sixth indices (noted hereafter as Z1),}
			{Z^\alpha}_{\beta\gamma}{^\mu}_{\nu\sigma} &=  -{Z^\alpha}_{\beta\sigma}{^\mu}_{\nu\gamma} \label{corr2formprop1}
			\intertext{The anti-symmetry in interchange of the index pairs (noted hereafter as Z2),}
			{Z^\alpha}_{\beta\gamma}{^\mu}_{\nu\sigma} &=  -{Z^\mu}_{\nu\gamma}{^\alpha}_{\beta\sigma} \label{corr2formprop2}
			\intertext{The algebraic cyclic identity (noted hereafter as Z3),}
			{Z^\alpha}_{\beta[\gamma}{^\mu}_{\nu\sigma]} &= 0 \label{corr2formprop3}
			\intertext{The equi-affinity property (noted hereafter as Z4),}
			{Z^\epsilon}_{\epsilon\gamma}{^\mu}_{\nu\sigma} &= 0 \label{corr2formprop4}
			\intertext{The differential constraint (noted hereafter as Z5),}	
			{Z^\alpha}_{\beta[\gamma}{^\mu}_{\underline{\nu}\sigma\|\lambda]} &= 0 \label{corr2formprop6}
			\intertext{The integrability condition (noted hereafter as Z6),}
			{Z^\epsilon}_{\beta[\gamma}{^\mu}_{\underline{\nu}\sigma}{M^\alpha}_{\underline{\epsilon}\lambda\rho]} - {Z^\alpha}_{\epsilon[\gamma}{^\mu}_{\underline{\nu}\sigma}{M^\epsilon}_{\underline{\beta}\lambda\rho]}& + {Z^\alpha}_{\beta[\gamma}{^\epsilon}_{\underline{\nu}\sigma}{M^\mu}_{\underline{\epsilon}\lambda\rho]} - {Z^\alpha}_{\beta[\gamma}{^\mu}_{\underline{\epsilon}\sigma}{M^\epsilon}_{\underline{\nu}\lambda\rho]} = 0 \label{corr2formprop7}
		\end{align}
		\noindent\textrm{The quadratic constraint (noted hereafter as Z7),}
		\begin{multline}\label{corr2formprop8}
			{Z^\delta}_{\beta[\gamma}{^\theta}_{\underline{\kappa}\pi}{Z^\alpha}_{\underline{\delta}\epsilon}{^\mu}_{\underline{\nu}\sigma]} + {Z^\delta}_{\beta[\gamma}{^\mu}_{\underline{\nu}\sigma}{Z^\theta}_{\underline{\kappa}\pi}{Z^\alpha}_{\underline{\delta}\epsilon]} + {Z^\alpha}_{\beta[\gamma}{^\delta}_{\underline{\nu}\sigma}{Z^\mu}_{\underline{\delta}\epsilon}{^\theta}_{\underline{\kappa}\pi]} + {Z^\alpha}_{\beta[\gamma}{^\mu}_{\underline{\delta}\epsilon}{Z^\theta}_{\underline{\kappa}\pi}{^\delta}_{\underline{\nu}\sigma]}\\ + {Z^\alpha}_{\beta[\gamma}{^\theta}_{\underline{\delta}\epsilon}{Z^\mu}_{\underline{\nu}\sigma}{^\delta}_{\underline{\kappa}\pi]} + {Z^\alpha}_{\beta[\gamma}{^\delta}_{\underline{\kappa}\pi}{Z^\theta}_{\underline{\delta}\epsilon}{^\mu}_{\underline{\nu}\sigma]} = 0 
		\end{multline}
	\end{subequations}
	
	\subsubsection{The Affine Deformation Tensor}
	The affine deformation tensor, through its definition, has the following field equations \cite{zala1,zala2,zala3,zala4,zala5,zala6,zala7},
	\begin{subequations}\label{affdefprop}
		\begin{equation}\label{affdefprop1}
			{A^\alpha}_{\beta[\rho\|\sigma]} - {A^\epsilon}_{\beta[\rho}{A^\alpha}_{\underline{\epsilon}\sigma]} = -\hf{Q^\alpha}_{\beta\rho\sigma}
		\end{equation}
		
		As a result of equation \eqref{corr2formprop6}, the field equations for the non-Riemannian macroscopic curvature tensor become $ {R^\alpha}_{\beta\rho\sigma\|\lambda} = 0 $. Using equation \eqref{curvdef2}, this takes the form,
		\begin{equation}\label{affdefprop2}
			{A^\alpha}_{\epsilon[\rho}{M^\epsilon}_{\underline{\beta}\sigma\lambda]} -{A^\alpha}_{\epsilon[\rho}{Q^\epsilon}_{\underline{\beta}\sigma\lambda]} - {A^\epsilon}_{\beta[\rho}{M^\alpha}_{\underline{\epsilon}\sigma\lambda]} + {A^\epsilon}_{\beta[\rho}{Q^\alpha}_{\underline{\epsilon}\sigma\lambda]}  = 0 
		\end{equation}
	\end{subequations}
	
	The macroscopic gravity equations \eqref{macroefe}, \eqref{corr2formprop} and \eqref{affdefprop} constitute a system of coupled first and second order non-linear partial differential equations for the connection correlation, affine deformation tensor and a (not yet assumed) macroscopic metric. Given a macroscopic metric, these equations can be solved to calculate components of the connection correlation and hence the additional terms in the mEFEs. Several such solutions have been found with interesting implications.
	
	\section{Solutions to the Macroscopic Einstein Field Equations} \label{mgsol}	
	
		It is worth recalling here that there are two common approaches to averaging in general relativity and cosmology.
		
		The first is what comes to ones' mind when considering averaging in a general context and similar to simple settings like in Newtonian gravity for example. That is, one thinks about taking a tensor or a scalar and apply to it a direct and explicit averaging operation. This also has been done in general relativity -- one can take a given microscope metric and explicitly calculate the average of this metric. This approach has it pros and cons but it is in general not trivial to find microscopic metrics or covariant averaging schemes to work with. For example, \cite{coleyss1,coleyss2} applied this direct approach to average spherically symmetric microscopic geometries. They worked with the volume preserving coordinates \cite{zala3} in which the averaging bi-vectors are simply the Kronecker delta and hence the process of averaging, in this case, is simplified.
		
		The second approach that we followed here is where one takes an ansatz on the macroscopic geometry and finds the back-reaction using the algebraic and differential constraints on the connection correlation. As mentioned earlier, when one uses a metric to describe the universe as a whole, we are intrinsically taking an ansatz on the macroscopic metric. Therefore, this approach follows the same way of model-building with the mEFEs as is done with just the EFEs. This can be viewed as an inverse approach to the problem and complementary to the other method described above. This approach has been widely used to analyse models within macroscopic gravity \cite{coley,hoogen2,hoogenss,clifton,tharake1,tharake2} and we have followed it in this work.
	
	The procedure to find the solutions to the mEFEs has been described extensively in the literature. Taking specific forms of $\bz$, one can solve the constraint equations Z1-Z7 to determine the connection correlation completely. The solution for an FLRW macroscopic metric was first presented in \cite{coley} and further elaborated upon in \cite{hoogen2}. A systematic approach to solving the MG equations has been presented in \cite{zala5,hoogen2,tharake1}. However, it is useful describe it here briefly. In the next two sections, we work following the same path as in \cite{hoogen2}.
	
	The connection correlation is a six rank tensor with 4096 components. These components are constrained through its various algebraic and differential properties discussed in the previous section. A step by step approach to solving for the components of this tensor is as follows,
	\begin{enumerate}[nosep]
		\item[1.] Properties Z1 and Z2 (equations \eqref{corr2formprop1} and \eqref{corr2formprop2}) reduce the number of independent components in $ \bz $ to 720.
		\item[2.] Property Z3 (equation \eqref{corr2formprop3}) reduces this number to 470.
		\item[3.] Property Z4 (equation \eqref{corr2formprop4}) further reduces this to 396.
	\end{enumerate}
	Now, we are left with is properties Z5, Z6, Z7. It was found in \cite{hoogen2} that this property is identically satisfied if we assume the `electric part' of $\bz$ to be zero. That is,
	\begin{equation}
		{{{ZE^\alpha}_\beta}^\gamma}_{\delta\mu} = {{{Z^\alpha}_{\beta\mu}}^\gamma}_{\delta\nu}u^\nu = 0
	\end{equation}
	where, $ u^\nu $ is the time like unit vector field admitted by the macroscopic geometry.
	
	\begin{enumerate}[nosep]
		\item[4.] Setting the electric part of $\bz$ to zero further reduces its independent components to 121.
	\end{enumerate}
	
	The steps described above do not depend on the macroscopic geometry. In other words, for any given macroscopic geometry, the connection correlation will have, at most, 121 independent components. Now, we are left with properties Z6 and Z7. Solving these requires a macroscopic metric and specifying the functional form of $\bz$. We do this, first here with an exact-FLRW metric and then, a linearly perturbed FLRW metric in the next sections.
	
	The line element for the FLRW space-time is given by,
	\begin{equation}\label{exactflrw}
		\df s^2 = a^2(\tau)\left(-\df \tau^2 + \delta_{ij}\df x^i \df x^j\right)
	\end{equation}
	where, $\tau \equiv \int \frac{\df t}{a(t)}$ is the conformal time.
	
	Starting with the 121 independent components, constraint Z6 reduces this number to 69. Applying constraint Z5 makes all of these 69 components independent of time. Further, taking $\bz$ to be invariant under the $\mathcal{G}_6$ group of motions as the metric, reduces the number of independent components to 3. These 3 components are arbitrary constants.  We will label these variables as $ b_1 $, $ b_2 $ and $ b_3 $. The MG correction term has only one of these variables and is found to be,
	\begin{subequations}\label{mgcorflrw}
		\begin{align}
			{C^0}_0 &= \frac{\mathcal{B}}{a^2(\tau)} \\
			{C^0}_i &= 0 = {C^i}_0 \\
			{C^i}_j &= \frac{1}{3}\frac{\mathcal{B}}{a^2(\tau)} \delta^i_j 
		\end{align}
	\end{subequations}
	In the matrix form, we have,
	\begin{equation}\label{mgcorflrwmatrix}
		{C^\epsilon}_\gamma = \frac{\mathcal{B}}{a(\tau)^2}\begin{bmatrix}
			&1 &0 &0 &0&\\
			&0 &\frac{1}{3} &0 &0&\\
			&0 &0 &\frac{1}{3} &0&\\
			&0 &0 &0 &\frac{1}{3}&
		\end{bmatrix}
	\end{equation}	
	where, $ \mathcal{B} = 15 b_2 $. The effect of the MG correction term is that of an additional spatial curvature (as previously found in \cite{coley,hoogen2,zala5,clifton,tharake1}) in the sense that it evolves as $a^{-2}$ in the Friedmann equations. 	
	
	Considering the average of the microscopic matter distribution to be a perfect fluid, we have,
	\begin{equation}\label{emtensor}
		^{{\rm (fluid)}}\la {t^\epsilon}_\gamma \ra = (\bar{\rho} + \bar{p}) \bar{u}^\epsilon \bar{u}_\gamma + \bar{p} \delta^\epsilon_\gamma
	\end{equation}
	where, $\bar{\rho} \equiv \bar{\rho}(\tau)$ and $\bar{p} \equiv\bar{p}(\tau)$ are the energy density and pressure for the averaged matter-energy content, respectively; and $ \bar{u}~=~\left[\frac{1}{a},0,0,0\right] $ is the averaged 4-velocity of the fluid. Then, writing the mEFEs (equation \eqref{macroefe1}) explicitly gives us the modified Friedmann equations,	
	\begin{subequations}\label{fried}			
		\begin{align}
			3\mathcal{H}^2 &= 8\pi \bar{\rho} a^2 - \mathcal{B} \label{fried1} \\
			2 \mathcal{H}^\prime + \mathcal{H}^2 &= -8\pi \bar{p} a^2 - \frac{1}{3}\mathcal{B} \label{fried2}
		\end{align} 
	\end{subequations}
	where, $\mathcal{H} \equiv \frac{a^\prime}{a}$ and the `prime' represents differentiation with respect to the conformal time, $\tau$.
	
	\section{Solutions for an Almost-FLRW Macroscopic Metric} \label{almostflrw}
	We will now describe new results calculating the MG correction term with an almost-FLRW macroscopic geometry, given by a linearly perturbed spatially flat FLRW metric. The line element for such a geometry, in the so called (conformal) Newtonian gauge, is given by,
	\begin{equation}\label{pflrw}
		ds^2 = a^2(\tau)\left[-\left(1 + 2\Psi\right)\df \tau^2 + \left(1 - 2\Phi\right) \delta_{ij}\df x^i \df x^j \right]
	\end{equation}
	where, $ \Psi $ and $ \Phi $ are the gauge invariant variables characterising the first order scalar perturbations. It is worth noting here that we are using the metric in equation \eqref{pflrw} as an ansatz for the macroscopic metric. Analysing cosmological perturbations explicitly within the framework of macroscopic gravity is a non trivial exercise. The issue of gauge invariance becomes important and several assumptions are required to do cosmological perturbation theory consistently within MG \cite{para1,para2}. These issues are not applicable here since we are taking the perturbed FLRW geometry itself as the macroscopic geometry. That is, equation \eqref{pflrw} describes the geometry of the universe after being averaged. This is consistent with other analyses of cosmological perturbation theory with MG in the literature \cite{clifton,tharake1,tharake2}.
	
	To find the back-reaction within this macroscopic geometry, we take the components of connection correlation to be space-dependent i.e., $\bz\equiv~\bz(x,y,z)$. We solve the constraint equations Z1-Z7 to derive the form of the the connection correlation and hence the MG correction tensor. Solving Z1 to Z4 and taking the electric part of $\bz$ to be zero leaves us with 121 independent components (they do not depend on the macroscopic geometry). Then we apply constraints Z5 and Z6 assuming the macroscopic geometry to be the one in equation \eqref{pflrw}. We do this for multiple cases with various assumptions on the functional form of the perturbations. We will list the results below one by one.
	
	\subsection{Perturbations as Arbitrary Functions of Time}
	This is the case of spatially homogeneous and isotropic perturbations. This is the same as taking an exact-FLRW metric to be the macroscopic metric. Therefore, this is equivalent to taking $\Psi = 0 = \Phi$. Obviously, solving the constraint equations \eqref{corr2formprop} gives the same results as for the exact-FLRW case and the MG correction tensor is the same as that found in equation \eqref{mgcorflrw}, with a rescaled time and the scale factor. This is the case that was previously analysed in \cite{clifton}. We will go beyond this case in the next sections.
	
	\subsection{Perturbations as Arbitrary Functions of Time and One Space Coordinate} \label{arbpertonespace}
	In this case, the constraint equations need to be solved explicitly with a perturbed metric. We do this to find that constraint Z6 reduces the independent components from $ 121 $ to $ 3 $ while Z7 is trivially satisfied, and does not reduce the independent components further. We label these variables as $f_1,f_2,f_3$. For a given choice of space coordinate dependence of the perturbations, these variables are arbitrary functions of the two other spatial coordinates that the perturbations do not depend on. Only two of these three components are there in the MG correction term. Characterising the sub-cases here by a variable $N \in \{1,2,3\}$, our results for the MG correction term can be written as,
	\begin{subequations}\label{mgcor2}
		\begin{align}
			{C^0}_0 &= \frac{\mathcal{B}}{a^2(\tau)\left\{1-2\Phi(\tau,q_N)\right\}} \\
			{C^0}_i &= 0 = {C^i}_0 \\
			{C^i}_j &= \frac{\mathcal{B}}{a^2(\tau)\left\{1-2\Phi(\tau,q_N)\right\}} \delta^i_j \qquad &&\ i=N  \\
			{C^i}_j &= 0  &&\ i\neq N 
		\end{align}
	\end{subequations}
	In the matrix form, we have,
	\begin{equation}\label{mgcor2matrix}
		{C^\epsilon}_\gamma = \frac{\mathcal{B}}{a(\tau)^2\left\{1-2\Phi(\tau,q_N)\right\}}\begin{bmatrix}
			&1 &0 &0 &0&\\
			&0 & & & &\\
			&0 & &\delta^N_j & &\\
			&0 & & & &
		\end{bmatrix}
	\end{equation}	
	where, we have,
	\begin{equation} \label{sdeparb1br}
		\mathcal{B}\ ;\; q_N = \begin{cases}
			2 f_1(y,z) + 2 f_2(y,z)\ ;\; x \quad \  &N=1\\
			2 f_1(x,z) + 2 f_2(x,z)\ ;\; y \quad \  &N=2\\
			2 f_1(x,y) + 2 f_2(x,y)\ ;\; z \quad \  &N=3
		\end{cases}
	\end{equation}	 
	
	This is a very restrictive form of the perturbations but the only one (other than homogeneous perturbations) that gives an explicit solution for the back-reaction. In all other more general cases, we recover a trivial solution to the constraint equations due to limitations in computing other non-trivial solutions.
	
	\section{Solutions for Plane Wave Perturbations} \label{pwa}
	The results in the section above are novel, however, as has been found in other work \cite{clifton}, to find more general solutions with the perturbations as functions of time and two or more spatial coordinates, one needs to specify a functional form of the perturbations. Following the commonly used ansatz in cosmological perturbation theory \cite{mukha1,sasaki,mukha2,weinc,dodel}, we use plane wave perturbations, $\Phi\sim \phi(\tau) \ex^{\io k_jx_j}$, where, $ \phi(\tau) $ is an arbitrary function of time and $ k_j $ is the wave-vector $(k_1,k_2,k_3)$. Since we have only taken scalar perturbations in our macroscopic metric (equation \eqref{pflrw}), these plane waves will be acoustic waves, since the scalar perturbations correspond to matter density perturbations in the universe. Working with this ansatz, we analyse all the previously described cases. Now, in all these cases, constraint Z6 reduces the independent components in $\bz$ from $ 121 $ to $ 3 $ while Z7 is trivially satisfied, and does not reduce the independent components further. These components are arbitrary functions of space. We label these functions as $f_1,\ f_2,\ f_3$. Note that these are not necessarily the same functions as those in the previous sections. Only two of these three functions are present in the MG correction term which takes the form,
	\begin{subequations}\label{mgcorpwa3}
		\begin{align}
			{C^0}_0 &= \frac{\mathcal{B}_W\ K}{a^2(\tau)\left(1-2\Phi\right)}k_ik_i \\
			{C^i}_0 &= 0 = {C^0}_i \\
			{C^i}_j &=\frac{\mathcal{B}_W\ K}{a^2(\tau)\left(1-2\Phi\right)}k_i k_j 
		\end{align}
	\end{subequations} 
	In the matrix form, we have,
	\begin{equation}\label{mgcorpwa3matrix}
		{C^\epsilon}_\gamma = \frac{\mathcal{B}_W\ K}{a(\tau)^2(1 - 2\Phi)}
		\begin{bmatrix}
			&k_ik_i &0 &0 &0&\\
			&0 & & & &\\
			&0 & &k_ik_j & &\\
			&0 & & & &
		\end{bmatrix}
	\end{equation}
	where, $\mathcal{B}_W = 2 f_1(\tilde{v},\tilde{w}) + 2 f_2(\tilde{v},\tilde{w}) $, $ \tilde{v}~=~\frac{k_1y - k_2x}{k1}$, $\tilde{w}~=~\frac{k_1z - k_3x}{k1}  $, and,
	$K = \frac{k_jk_j}{k_1^2(k_1^2 + k_2^2)}$.
	
	When we consider only the real part of the plane wave, i.e, $ \Phi \sim \phi(\tau) \cos(k_ix_i) $, our results remain the same as in equation \eqref{mgcorpwa3}, with the perturbation term in the denominator given by a cosine function instead of an exponential. 
	
	\section{Macroscopic Einstein Field Equations for an Almost-FLRW Metric} \label{secpertmefe}
	In the previous two sections, we presented our results for the MG correction term for an almost-FLRW universe. Here, we present the full field equations for an almost-FLRW metric including this correction. The MG correction leads to an extra term in both the zeroth and first order equations. This modifies the dynamical evolution of background as well as perturbations. 
	
	The macroscopic metric is the one in equation \eqref{pflrw}. The average of microscopic energy-momentum tensor is that of a perturbed perfect fluid, given by,
	\begin{equation}\label{pertemtensor}
		^{{\rm (pfluid)}}\la {t^\epsilon}_\gamma \ra =  \{(\bar{\rho}+\delta\rho) + (\bar{p} + \delta p)\}u^\epsilon u_\gamma + (\bar{p}+\delta p)\delta^\epsilon_\gamma	
	\end{equation}
	where, $ \delta\rho \equiv \delta \rho(\tau,x,y,z) $ and $ \delta p \equiv \delta p(\tau,x,y,z) $ are the perturbations to the energy density and pressure. The 4-velocity of the perturbed fluid is given by, $u^\nu =~\frac{1}{a}\left[\left(1-\Psi\right), \delta u^i \right] $, where, $ \delta u^i $ is sometimes called the peculiar velocity. Note that the order of the perturbations in the energy-momentum tensor is same as the metric perturbations, i.e., $ \Psi \sim \Phi \sim \delta \rho \sim \delta p\sim\mathcal{O}(\epsilon)$ (say), where, $\epsilon << 1 $, is a small parameter.
	
	We present the mEFEs by calculating the Einstein tensor for the given metric, equating it to the right hand side in equation \eqref{macroefe} and then restricting all the terms to be only up to first (linear) order in perturbations. This gives the following equations,
	\begin{subequations}\label{pertmefe}
		\begin{align} 
			3\mathcal{H}^2 + 2\partial^i\partial_i \Phi - 6\mathcal{H}(\Phi^\prime + \mathcal{H}\Psi) &= 8\pi\bar{\rho}a^2 + 8\pi \delta\rho a^2
			- \Br - 2\Br\Phi\label{pertmefe1}\\
			\partial_i\left(\Phi^\prime + \mathcal{H}\Psi \right) &= 4\pi (\bar{\rho}+\bar{p})\delta u_i a^3 \label{pertmefe2}\\			
			\partial^i\partial_j (\Psi - \Phi) &= -{C^i}_j a^2\qquad\qquad\qquad i\neq j \label{pertmefe3}
		\end{align}
		\begin{multline}
			2\mathcal{H}^\prime + \mathcal{H}^2 - 2\Phi^{\prime\prime} - 4\mathcal{H}\Phi^\prime - 2\mathcal{H}\Psi^\prime - 2(2\mathcal{H}^\prime + \mathcal{H}^2)\Psi + \frac{2}{3}\partial^i\partial_i (\Psi-\Phi) \\= -8\pi \bar{p}a^2 - 8\pi\delta p a^2 - \frac{1}{3}\Br - \frac{2}{3}\Br\Phi \label{pertmefe4}
		\end{multline}
	\end{subequations}  
	where, $\Br = 2(f_1+f_2)$, and $f_1,f_2$ are the arbitrary functions of spatial coordinates arising in the MG correction term.
	
	Then, at first order, we have,
	\begin{subequations}\label{modpert}
		\begin{align} 
			\partial^i\partial_i \Phi - 3\mathcal{H}(\Phi^\prime + \mathcal{H}\Psi) &= 4\pi \delta\rho a^2 - \Br\Phi \label{modpert1}\\
			\partial_i\left(\Phi^\prime + \mathcal{H}\Psi \right) &= 4\pi (\bar{\rho}+\bar{p})\delta u_i a^3 \label{modpert2}\\			
			\partial^i\partial_j (\Psi - \Phi) &= -{C^i}_j a^2 \label{modpert3}\\ 
			\Phi^{\prime\prime} + 2\mathcal{H}\Phi^\prime + \mathcal{H}\Psi^\prime + (2\mathcal{H}^\prime + \mathcal{H}^2)\Psi &- \frac{1}{3}\partial^i\partial_i (\Phi-\Psi) =  4\pi\delta p a^2 + \frac{1}{3}\Br\Phi \label{modpert4}
		\end{align}
	\end{subequations} 	
	The arbitrary functions involved in the almost-FLRW solutions can be assumed to be equal to the constants involved in the exact-FLRW solutions to get the above first order equations. The full set of field equations is used so that the terms at the zeroth and the first orders separate out as they should. Equations \eqref{modpert} give us the dynamical evolution equations for the scalar part of the gauge invariant metric perturbations. The equations above tell us that the geometric correction in MG affects both the background evolution as well as the perturbations around this background.
	
	In an expanding universe, the second and third terms on the left hand side of equation \eqref{modpert1} can be neglected on sub-Hubble scales \cite{mukha1,sasaki,mukha2} (they are suppressed by a factor of $\sim \lambda H$), and hence, it again reduces to the Poisson equation (with a modified source). In the usual cosmological perturbation theory with GR, the right hand side of equation \eqref{modpert3} is equal to zero in the absence of anisotropic stress in the fluid (equation \eqref{pertemtensor}). This is interpreted as the two potentials being equal $(\Phi = \Psi)$ at late times, that is, the `slip parameter' $ (\Phi - \Psi) $ being zero. However, in MG, even with zero anisotropic stress, a non-zero slip parameter could exist, solely due to the correction tensor. In other words, one of the effects averaging in cosmology could lead to is that of a non-zero slip parameter (an effective non-zero anisotropic stress) even when the anisotropic stress in the energy-momentum tensor is taken to be zero. In our analysis here, we find that the cases with plane wave perturbations depending on two or more spatial coordinates lead to this effect. For all the other cases, the slip parameter is still zero. 	
	
	\section{Conclusion}\label{conclusion}
	In this paper, we analysed almost-FLRW cosmological models within the framework of macroscopic gravity. We noted that the effects due to averaging appear, not only when considering the largest scales (where the geometry is exact-FLRW), but also at slightly smaller length scales (where the geometry is almost-FLRW). When going from small lumpy scales to the largest smooth scales, one can consider that the geometry at the intermediate scales is represented by some particular corresponding metrics. Then, at a given scale, one would have different values of the back-reaction related to the corresponding metric. The back-reaction at every such scale would be an integrated effect of all the `coarse graining' employed to reach the level of smoothness represented by the metric at this scale.
	
	We took the macroscopic metric to be almost-FLRW by using an FLRW metric and linear perturbations around it. We computed the macroscopic gravity field equations including the back-reaction for such a metric and restricted all the terms to be up to first order in perturbations. In doing so, we analysed several cases with various functional forms and spatial/temporal dependencies of the perturbations and found the macroscopic gravity correction term. These functional dependencies correspond to different coordinate ansatzes. Therefore, it should be noted that explicit coordinate ansatzes might lead to some coordinate specific effects. In particular, in case of the plane wave ansatz, the wave vector can always be chosen to be oriented such that the plane wave is a function of only one spatial coordinate. Hence, effects of averaging in this case might be explicit only in specific coordinate choice and thus is of a limited scope, but nevertheless, was necessary in order to be able to analyse (macroscopic) metric perturbations since the mathematical complexity of MG makes it difficult to do so in general terms. In sum, using this approach, we were able to show that the use of special functional forms of perturbations can lead to general solutions with back-reaction terms not only at the background level but also at the perturbed level confirming explicitly the non-linearity of the averaging procedure resulting from its non-commutativity with the construction of the Einstein tensor. 
	
	\section*{Acknowledgements}
		We would like to thank Robert J. van den Hoogen, Timothy Clifton and Alan Coley for useful comments. The results in this paper were obtained by modification and expansion of a code originally written by Tharake Wijenayake and Mustapha Ishak \cite{tharake1}, using the computer algebra software \href{https://www.maplesoft.com}{Maple} and the openly available package \href{https://github.com/grtensor/grtensor}{GRTensor}. MI acknowledges that this material is based upon work supported in part by the Department of Energy, Office of Science, under Award Number DE-SC0022184.
	
	\setlength{\bibsep}{1pt}
	\bibliographystyle{unsrtnat}
	\bibliography{pflrwinmg}
\end{document}